\pdfoutput=1
\documentclass[a4paper,compress]{elsarticle}

\usepackage{amsmath,amssymb}
\usepackage{appendix}
\usepackage{graphics}
\usepackage[utf8x]{inputenc}
\usepackage[nooneline]{subfigure}
\usepackage[textsize=footnotesize,textwidth=30mm]{todonotes}
\usepackage{xspace}
\usepackage{hyperref}
\usepackage{multirow}

\biboptions{numbers, sort&compress}
\journal{Computer Physics Communications}

\newlength{\figwidth}
\newcommand{\cmistark}{\textsc{CMIstark}\xspace}
\newcommand{\degree}[1]{\xspace\ensuremath{^\circ}#1}%
\newcommand{\eg}{e.\,g.}%
\newcommand{\esp}{esp.\ }%
\newcommand{\field}{\ensuremath{\varepsilon}\xspace}%
\newcommand{\ie}{i.\,e.}%
\newcommand{\mueff}{\ensuremath{\mu_\text{eff}}\xspace}%
\newcommand{\op}[1]{\ensuremath{\mathbf{#1}}}%

\begin{document}
\begin{frontmatter}
   \title{\cmistark: Python package for the Stark-effect calculation and symmetry classification of
      linear, symmetric and asymmetric top wavefunctions in dc electric fields}%
   \author[cfel]{Yuan-Pin Chang}%
   \author[fhi]{Frank Filsinger\fnref{bruker}}%
   \fntext[bruker]{Present address: Bruker AXS GmbH, Karlsruhe, Germany}%
   \author[ras]{Boris G.\ Sartakov}%
   \author[cfel,uhh,cui]{Jochen Küpper\corref{cor1}}%
   \cortext[cor1]{Corresponding
      author}\ead{jochen.kuepper@cfel.de}\ead[url]{http://desy.cfel.de/cid/cmi}%
   \address[cfel]{Center for Free-Electron Laser Science, DESY, Notkestrasse 85, 22607 Hamburg, Germany}%
   \address[fhi]{Fritz-Haber-Institut der MPG, Faradayweg 4-6, 14195 Berlin, Germany}%
   \address[ras]{General Physics Institute RAS, Vavilov str. 38, 119991, Moscow, Russia}%
   \address[uhh]{\mbox{Department of Physics, University of Hamburg, Luruper Chausse 149, 22761 Hamburg, Germany}}%
   \address[cui]{\mbox{The Hamburg Center for Ultrafast Imaging, Luruper Chaussee 149, 22761 Hamburg, Germany}}%
   \begin{abstract}
      The Controlled Molecule Imaging group (CMI) at the Center for Free Electron Laser Science
      (CFEL) has developed the \cmistark\ software to calculate, view, and analyze the energy levels
      of adiabatic Stark energy curves of linear, symmetric top and asymmetric top molecules. The
      program exploits the symmetry of the Hamiltonian to generate fully labeled adiabatic Stark
      energy curves.

      \cmistark\ is written in Python and easily extendable, while the core numerical calculations
      make use of machine optimized BLAS and LAPACK routines. Calculated energies are stored in HDF5
      files for convenient access and programs to extract ASCII data or to generate graphical plots
      are provided.
   \end{abstract}
   \begin{keyword}
      molecular rotation \sep linear top molecule \sep symmetric top molecule \sep asymmetric top
      molecule \sep electric field \sep Stark effect
   \end{keyword}
\end{frontmatter}

\section{Program summary}
\label{sec:program-summary}
\paragraph{Program title:} \cmistark
\paragraph{Catalogue identifier:} (to be added in production)
\paragraph{Program summary URL:} (to be added in production)
\paragraph{Program obtainable from:} CPC Program Library
\paragraph{Licensing provisions:} GNU General Public License version 3 or later with amendments. See code for details.
\paragraph{No. of lines in distributed program, including test data, etc.:} 3394
\paragraph{No. of bytes in distributed program, including test data, etc.:} 61962
\paragraph{Distribution format:} tar.gz
\paragraph{Programming language:} Python (version 2.6.x, 2.7.x)
\paragraph{Computer:} Any Macintosh, PC, or Linux/UNIX workstations with a modern Python distribution
\paragraph{Operating system:} Tested on Mac OS X and a variety of Linux distributions
\paragraph{RAM:} 2 GB for typical calculations
\paragraph{Classification:} Atomic and Molecular Physics, Physical Chemistry and Chemistry Physics
\paragraph{External routines:} Python packages numpy and scipy; utilizes (optimized) LAPACK and BLAS
through scipy. All packages available under open-source licenses.
\paragraph{Nature of problem:} Calculation of the Stark effect of asymetric top molecules in
arbitrarily strong dc electric fields in a correct symmetry classification and using correct
labeling of the adiabatic Stark curves.
\paragraph{Solution method:} We set up the full $M$ matrices of the quantum-mechanical Hamiltonian
in the basis set of symmetric top wavefunctions and, subsequently, Wang transform the Hamiltonian
matrix. We separate, as far as possible, the sub-matrices according to the remaining symmetry, and
then diagonalize the individual blocks. This application of the symmetry consideration to the
Hamiltonian allows an adiabatic correlation of the asymmetric top eigenstates in the dc electric
field to the field-free eigenstates. This directly yields correct adiabatic state labels and,
correspondingly, adiabatic Stark energy curves.
\paragraph{Restrictions:} The maximum value of $J$ is limited by the available main memory. A modern
desktop computer with 16 GB of main memory allows for calculations including all $J$s up to a values
larger than $100$ even for the most complex cases of asymmetric tops.
\paragraph{Additional comments:}
\paragraph{Running time:} Typically 1~s--1~week on a single CPU or equivalent on multi-CPU systems
(depending greatly on system size and RAM); parallelization through BLAS/LAPACK. For instance,
calculating all energies up to $J=25$ of indole (\emph{vide infra}) for one field strength takes
1~CPU-s on a current iMac.

\section{Introduction}
\label{sec:introduction}
Over the last decade, the manipulation of the motion of molecules using electric fields has been
revitalized~\cite{Meerakker:CR112:2012, Filsinger:PCCP13:2076,Schnell:ACIE48:6010, Bell:MP107:99,
   Kuepper:FD142:155}. Exploiting the Stark effect, large asymmetric-top polar molecules have been
deflected~\cite{Holmegaard:PRL102:023001}, focused~\cite{Filsinger:PRL100:133003}, and
decelerated~\cite{Wohlfart:PRA77:031404}. These techniques can be used to spatially separate neutral
molecules according to their quantum states~\cite{Nielsen:PCCP13:18971}, structural
isomers~\cite{Filsinger:PRL100:133003, Filsinger:ACIE48:6900}, and cluster
sizes~\cite{Trippel:PRA86:033202}. These techniques promise advanced applications of well-defined
samples of complex molecules in various research fields, \eg, modern
spectroscopies~\cite{Dian:Science320:924, Hennies:PRL104:193002} or the direct imaging of structural
and chemical dynamics~\cite{Filsinger:PCCP13:2076, Sciaini:RPP74:096101, Barty:ARPC64:415,
   Kuepper:LCLSdibn:inprep}. However, successful implementation of these methods requires a thorough
theoretical understanding of the molecule-field interaction for the involved molecular quantum
states. Here we provide a well-tested and optimized program package for the calculation and labeling
of so called Stark curves, \ie, the energies of molecules as a function of electric field strength,
for general use. This software package will benefit the advance of those forthcoming applications,
\esp also for complex molecules. Moreover, it allows non-specialists and newcomers to the field to
concentrate on their envisioned applications of controlled molecules.

The code presented here is designed to calculate eigenenergies of very cold (on the order of a few
Kelvin) ensembles of polar molecules in the presence of external electrostatic fields. The
interaction of the molecular dipole moment with the dc electric field changes the internal energy,
and this is called Stark effect. To quantify this behavior, the eigenvalue problem of the
Hamiltonian is solved. \cmistark does this calculation in terms of numerically diagonalizing the
corresponding Hamiltonian matrix. An efficient method of diagonalizing the matrix, exploiting
underlying physics phenomena, is employed. Moreover, a correct method of correlating eigenvalues to
quantum states, \ie, labeling the calculated energies for all field strengths, is also required for
further use in order to predict or simulate and analyze control experiments.

The software package is named \cmistark. It is developed and maintained by the Controlled Molecule
Imaging (CMI) group at the Center for Free Electron Laser Science (CFEL), DESY, in Hamburg, based on
earlier work by some of the authors at the Fritz Haber Institut of the MPG in Berlin.

\section{Description}
\label{sec:description}
Stark energies are obtained by setting up and diagonalizing the Hamiltonian matrix for a given
electric field strength. The matrix elements can be obtained analytically (\emph{vide infra}) and
the resulting matrix is diagonalized numerically to obtain its eigenvalues, corresponding to the
energies of the molecular states. First, the matrix is block-diagonalized as far as possible using
symmetry considerations in order to correctly assign quantum numbers to eigenvalues. The
block-diagonalization also significantly reduces the overall computation time, which is dominated by
the diagonalization. The resulting blocks are diagonalized using LAPACK's dsyevr or zheevr
subroutines for real and complex matrices, respectively. The following overview section will provide
a brief review of the main concepts of the above approach.

\subsection{Overview}
\label{sec:overview}
The quantum-mechanical energy of a molecule, $E$, can be obtained by solving the Schrödinger
equation
\begin{equation}
   H\Psi=E\Psi.
   \label{eq:sch_eq}
\end{equation}
Neglecting translation, $H$ denotes the Hamiltonian operator in the center-of-mass frame and $\Psi$
is the wavefunction. For a rigid rotor and neglecting nuclear hyperfine-structure effects, the
Hamiltonian can be expressed in terms of components of the total angular moment operator $\op{J}$
about the principal axes ($a,b,c$), \ie, $\op{J}_a,\op{J}_b,\op{J}_c$~\cite{Gordy:MWMolSpec,
   Zare:AngularMomentum}:
\begin{equation}
   H_{\text{rigid}}%
   = \hbar^2(\frac{\op{J}_a^2}{2I_a}+\frac{\op{J}_b^2}{2I_b}+\frac{\op{J}_c^2}{2I_c})%
   = h(A\op{J}_a^2+B\op{J}_b^2+C\op{J}_c^2),
   \label{eq:ham-rot}
\end{equation}
where $h$ is Planck's constant, $\hbar=h/2\pi$, and $I_a$, $I_b$ and $I_c$ are three principal
moments of inertia of the rotor. By convention, the principal axes of inertia ($a,b,c$) are labeled
such that $I_a\le{}I_b\le{}I_c$. Note that, in the program, instead of moments of inertia we use
rotational constants, $A,B,C$, which in units of Hertz (Hz) are~\cite{Gordy:MWMolSpec,
   Zare:AngularMomentum}:
\begin{equation}
   A=\frac{h}{8\pi^2I_a},\quad B = \frac{h}{8\pi ^2I_b},\quad C=\frac{h}{8\pi ^2I_c}.
   \label{eq:rot-const}
\end{equation}
Molecular rotors are classified in terms of the magnitudes of their inertial moments, or rotational
constants, as shown in \autoref{tab:rotor}.
\begin{table}[b]
   \centering
   \begin{tabular}{ccl}
      moments of inertia & rotational constant & rotor type \\
      \hline
      $I_a=0;I_b=I_c$ & $A=\infty,B=C$ & linear top \\
      $I_a=I_b=I_c$ & $A=B=C$ & spherical top \\
      $I_a<I_b=I_c$ & $A>B=C$ & prolate symmetric top \\
      $I_a=I_b<I_c$ & $A=B>C$ & oblate symmetric top \\
      $I_a\neq I_b\neq I_c$ & $A\neq B\neq C$ & asymmetric top \\
   \end{tabular}
   \caption{Types of rotors defined through their inertial parameters.}
   \label{tab:rotor}
\end{table}%
Several quantum numbers are used to denote zero-field wavefunctions and energies of the rotational
states of molecules~\cite{Gordy:MWMolSpec, Zare:AngularMomentum}. The Schrödinger equation of
$H_\text{rigid}$, \eqref{eq:sch_eq} and \eqref{eq:ham-rot}, of linear rotors and symmetric tops in
free space (see \autoref{tab:rotor}) can be solved analytically, and their eigenfunctions of
$H_\text{rigid}$ are expressed as spherical harmonics $|J,M\rangle$ and Wigner $D$ matrices
$|J,K,M\rangle$, respectively~\cite{Gordy:MWMolSpec, Zare:AngularMomentum}. $J$ represents the
quantum number of total angular momentum $\op{J}$, $K$ characterizes the projection of $\op{J}$ onto
the symmetry axis of the symmetric top, and $M$ is the quantum number characterizing the projection
of $\op{J}$ onto a space fixed $Z$-axis. For asymmetric tops $K$ is not a good quantum number and
the Schrödinger equation of $H_\text{rigid}$ cannot generally be solved analytically. A numerical
calculation uses symmetric top wavefunctions $\left|J,K,M\right>$ as a basis set for obtaining
asymmetric top eigenfunctions $\left|J_{K_aK_c}M\right>$. Here, the quantum number $J$ and two
pseudo quantum numbers $K_a,K_c$ specify the zero-field rotational states. Finally, we only focus on
closed shell molecules, \ie, molecules which do not have unpaired electrons. Therefore, the values
of $J$, $K$ and $M$ are integer.

However, a real molecular system is not rigid. It is assumed that all non-rigidity under the
experimental conditions (on the order of 1\,K) can be described by a Hamiltonian representing
centrifugal distortion, $H_\text{d}$, with corresponding centrifugal distortion
constants~\cite{Gordy:MWMolSpec}. The Hamiltonian for such a nonrigid rotor is thus written as:
$H_\text{rot} = H_\text{rigid} + H_\text{d}$. The further details of $H_\text{d}$ for each type of
rotor are described in the next sections. In all cases we have implemented the lowest order quartic
centrifugal distortion terms. Higher order terms can easily be added if necessary.

The Stark effect of a polar molecule in a dc electric field is dominated by the interaction
$\vec{\mu}\cdot\vec{\field}$ of the molecule's dipole moment $\vec{\mu}$ with the field
$\vec{\field}$. While higher-order effects become relevant in strong field, they can still be
neglected in our case. For instance, the permanent dipole moment of benzonitriles ground state leads
to an energy shift of 300~GHz at 200~kV/cm, but the corresponding effect due to the polarizability
of the very similar non-polar molecule benzene is only 50~MHz~\cite{Okruss:JCP110:10393}, \ie,
almost four orders of magnitude smaller.

The dipole interaction with the electric field is described by the following contribution to the
Hamiltonian:
\begin{equation}
   H_\text{Stark}=-\field\sum_{g=x,y,z}{\mu_g\phi_{Z_g}},
\label{eq:Hstark}
\end{equation}
where $x,y,z$ represent a molecule-fixed coordinate system, $\mu_g$ represent the dipole moment
components along the molecule-fixed axes $x,y,z$, and $\phi_{Z_g}$ are the direction cosines of the
$x,y,z$ axes with reference to the space-fixed $X,Y,Z$-axes. $Z$ is oriented along the
electric-field direction. In the program, the principle axis system $(a,b,c)$ is identified with the
molecule-fixed system $(x,y,z)$ in representation $I^r (x=b,y=c,z=a)$~\cite{Gordy:MWMolSpec,
   Zare:AngularMomentum}. Note that this definition has the advantage that the Stark Hamiltonian
does not mix states with different values of $K$ if the dipole moment is parallel to the molecular
$a$ axis. The rotational Hamiltonian in the field \field can thus be written as:
$H_\text{rot,\,\field} = H_\text{rot} + H_\text{Stark}$.

In the program, the Schrödinger equation of the Hamiltonian in the field, $H_\text{rot,\,\field}$,
is solved numerically. The corresponding Hamiltonian matrix and the strategy of its diagonalization
are described in following sections for each type of rotor. Finally, the program assigns the
calculated rotational energies in the field to ``adiabatic quantum numbers'', \ie, to the
adiabatically corresponding field-free rotor states~\cite{Filsinger:JCP131:064309}. To ensure
correct assignments, a symmetry classification of $H_\text{rot,\,\field}$ and quantum states
according the electric field symmetry group~\cite{Watson:CJP53:2210, Bunker:MolecularSymmetry} is
required. In addition to the separation of $M$ and, for the symmetric top, $K$, this is achieved
through an appropriate unitary transformation of $H_\text{rot,\,\field}$ following Wang's
method~\cite{Gordy:MWMolSpec, Wang:PR34:243}.

\subsection{Linear top}
\label{sec:linear-top}
In a linear polyatomic molecule, the moment of inertia about the principal axis $a$ is zero whereas
the two other moments of inertia along axes $b$ and $c$ are equal: $I_b=I_c=I$. The centrifugal
distortion Hamiltanion $H_\text{d}$ takes the form
\begin{equation}
   H_\text{d} = -hD\op{J}^4
\end{equation}
where $D$ (Hz) is a centrifugal distortion constant. The first-order perturbation energy, which is
the expectation value of $H_\text{d}$ over field-free linear rotor wavefunctions, $|J,M\rangle$, is
included in the Hamiltonian matrix. The dipole moment of a linear molecule is along its symmetry
axis $z$, \ie, $\mu_z=\mu$ and $\mu_x=\mu_y=0$. Thus the Stark Hamiltonian simply becomes
\begin{equation}
   H_\text{Stark} = -\mu\field\phi_{Z_z}
\label{eq:stark_lin}
\end{equation}
The non-zero matrix elements for $H_\text{rot}$ and $H_\text{Stark}$ in the basis of linear top
wavefunctions $|J,M\rangle$, are provided in \ref{app:lin_rot}. The Hamiltonian matrix is
diagonalized directly without any further simplification.

\subsection{Symmetric top}
\label{sec:symmetric-top}
A molecule in which two of the principal moments of inertia are equal is a symmetric-top rotor, such
as a prolate top ($I_a<I_b=I_c$) and an oblate top ($I_a=I_b<I_c$). The figure axis of the
molecule,\footnote{Here we consider axially symmetric molecules, ignoring molecules which
   accidentally have an equivalent tensor of inertia.} which is parallel to the dipole moment, must
lie along the special principal axis of inertia, \ie, along the $a$ axis for a prolate top and the
$c$ axis for an oblate top. The non-rigidity of a symmetric top is taken into account by including
the first order perturbation energy of the corresponding centrifugal distortion Hamiltonian
($H_\text{d}$)~\cite{Gordy:MWMolSpec} into the Hamiltonian matrix. The Stark Hamiltonian
$H_\text{Stark}$ of symmetric tops is the same as that for linear rotors, as shown in
\autoref{eq:stark_lin}. The matrix elements for $H_\text{rot}$ and $H_\text{Stark}$ in the basis of
symmetric top wavefunctions, $|J,K,M\rangle$, are listed in \ref{app:sym_rot}. Finally, the strategy
for diagonalizing symmetric and asymmetric top Hamiltonian matrices is the same, and is described in
the next section. Moreover, in the case of the symmetric top, $K$ is a good quantum number and an
additional factorization into separate $K$ blocks is possible.

\subsection{Asymmetric top}
\label{sec:asymmetric-top}
\begin{figure}
  \centering
  \includegraphics[width=0.5\linewidth]{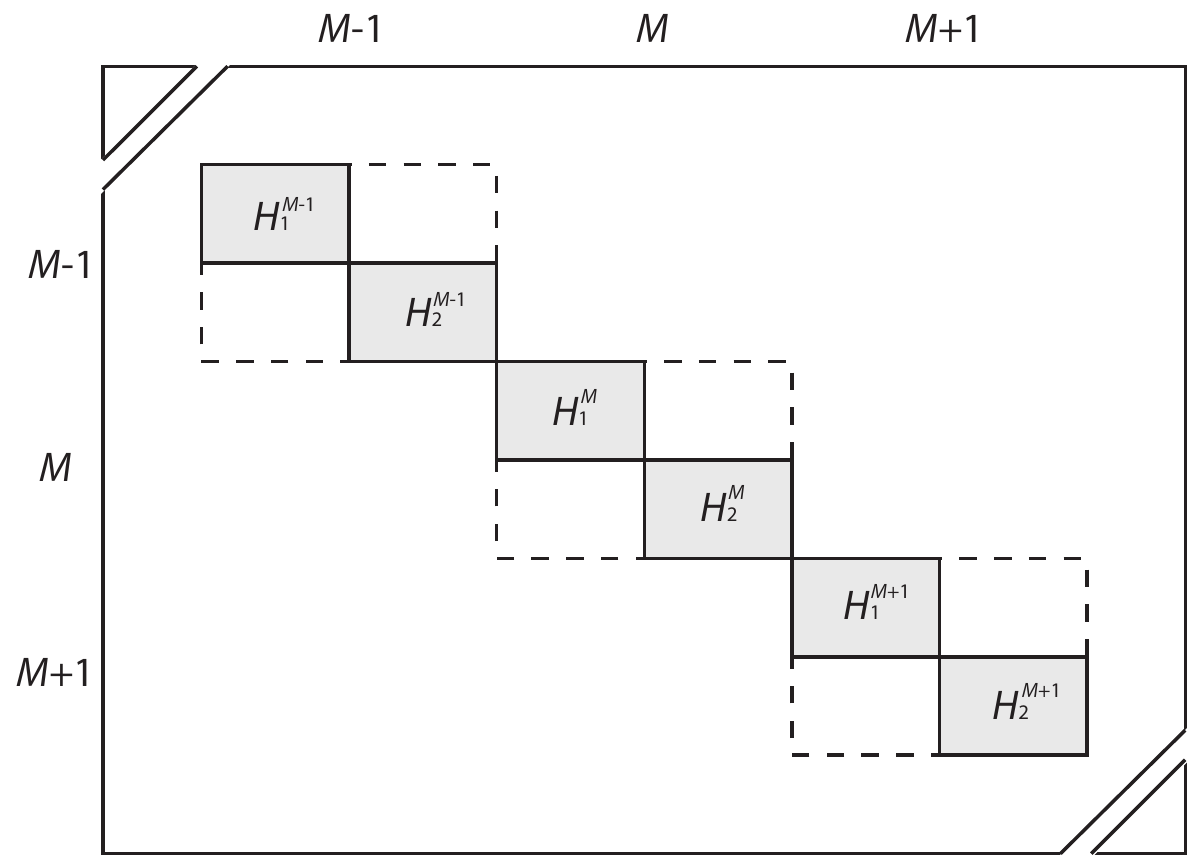}
  \caption{Schematic illustration of the factoring of the Hamiltonian matrix into blocks, labeled by
     $H_i^M$, in terms of $M$ (depicted by dashed lines) and symmetry. Nonzero elements are only
     present in the shaded blocks. The number of shaded blocks in each $M$ block depends on the
     molecular properties, \ie, the symmetry and the dipole moment direction; see text for details.}
  \label{fig:matrix-layout}
\end{figure}

An asymmetric-top molecule has three non-zero and non-equal principal moments of inertia. As
mentioned before, its Schrödinger equation even in the field-free case has no trivial analytical
solution for general $J$, and field-free symmetric top wavefunctions, $\left|J,K,M\right>$, are used
as the basis set for the Hamiltonian matrix $H_\text{rot,\,\field}$~\cite{Gordy:MWMolSpec}. All
nonzero matrix elements of $H_\text{rot,\,\field}$ in this basis set are listed in
\ref{app:asym_rot}. Note that, in this Hamiltonian matrix $H_\text{rot,\,\field}$, there are no
off-diagonal matrix elements in $M$, because $M$ is still a good quantum number in the field. Thus,
the blocks of each value of $M$ in the matrix, as shown in \autoref{fig:matrix-layout}, can be
diagonalized separately. As $K$ in the asymmetric top case and $J$ in the non-zero field case are
not good quantum numbers the set of basis functions for the block must cover a wide enough range of
$K$ and $J$ to ensure the accuracy of the numerical solution of the eigenvalue problem.

A further simplification of the matrix can be obtained by considering the symmetry properties of the
Hamiltonian. As mentioned before, the symmetry classification is required in order to distinguish
avoided crossings (between curves with the same symmetry) and real crossings (between curves with
different symmetries) and to assign the energy levels correctly. The field-free Hamiltonian operator
($H_\text{rigid}$ and $H_\text{rot}$) belongs to a symmetry group called Fourgroup and it is
designated by $V(a,b,c)$ (see \ref{app:four_group} for a detailed introduction). However, symmetric
top wavefunctions, which are the natural basis set, do not belong to the Fourgroup. A transformation
to a symmetrized basis is provided by the Wang transformation of the Hamiltonian
matrix~\cite{Gordy:MWMolSpec}:
\begin{equation}
   H_\text{Wang} = \widetilde{X}HX = \sum_iH_i
\end{equation}
where $X$ denotes the Wang transformation matrix and $H_i$ denotes a sub-matrix for each symmetry
species $i$. Thus, the Hamiltonian is expressed in a basis of linear combinations of symmetric top
wavefunctions which obey the Fourgroup symmetry~\cite{Gordy:MWMolSpec}. For a field-free asymmetric
top, its Wang transformed Hamiltonian matrix, \ie, $H_\text{Wang,\,rot}$, can be factorized into four
sub-matrices in terms of the Fourgroup symmetry species, as described in \ref{app:four_group}.

When an external field is applied, the number of sub-matrices in the Hamiltonian matrix,
$H_\text{Wang,\,rot,\,\field}$, usually reduces depending on the dipole moment direction in the
molecule and on the values of $M$. As described in \ref{app:Wang_trans} and \ref{app:sym_field}, if
the molecule's dipole moment is parallel to one principal axis of inertia,
$H_\text{Wang,\,rot,\,\field}$ can be factorized into two sub-matrices for $M\neq0$, and four for
the special case $M=0$. If the molecule's dipole moment is not parallel to any principal axis of
inertia, no factorization of $H_\text{Wang,\,rot,\,\field}$ is possible for $M\neq0$. For $M=0$ a
factorization into two blocks is still possible if one dipole moment component
$\mu_\alpha(\alpha=a,b,c)$ is zero. In any case, the above block diagonalization ensures that all
eigenstates obtained from the diagonalization of each sub-matrix $H_i^M$ (see
\autoref{fig:matrix-layout}) belong to a same symmetry. This means that all crossings between
eigenstates of each sub-matrix are avoided and that the energy order of these states remains the
same adiabatically, at any field strength. As a result, we can sort the resulting states of each
$H_i^M$ by energy and assign quantum number labels in the same order as for energy-sorted field-free
states of the same symmetry. This yields a correct assignment of ``adiabatic quantum number
labels'', $\tilde{J}_{\tilde{K}_a\tilde{K}_c}\tilde{M}$, to rotational states in the
field~\cite{Filsinger:JCP131:064309}.

\subsection{Results}\label{sec:results}
\begin{figure}
   \centering%
   \subfigure[OCS]{\includegraphics[width=\linewidth]{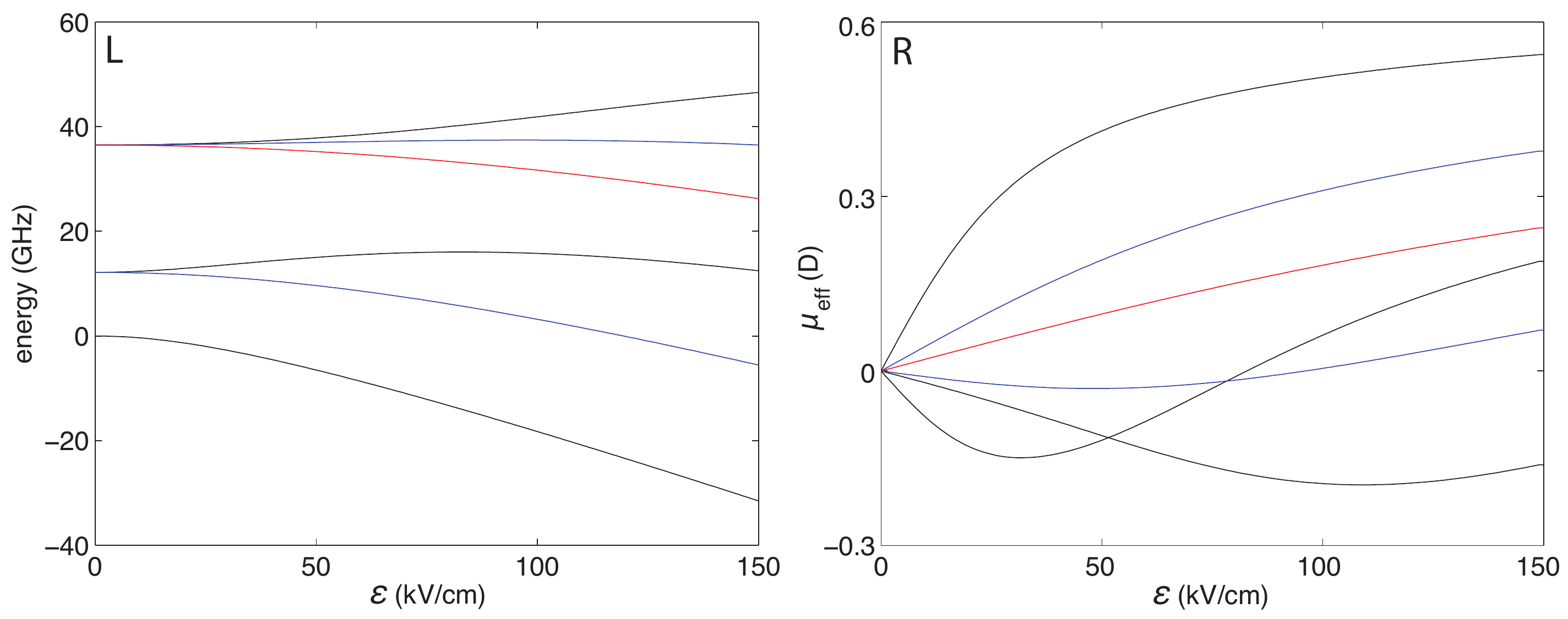}\label{fig:stark-curve-1}}\\
   \subfigure[iodomethane]{\includegraphics[width=\linewidth]{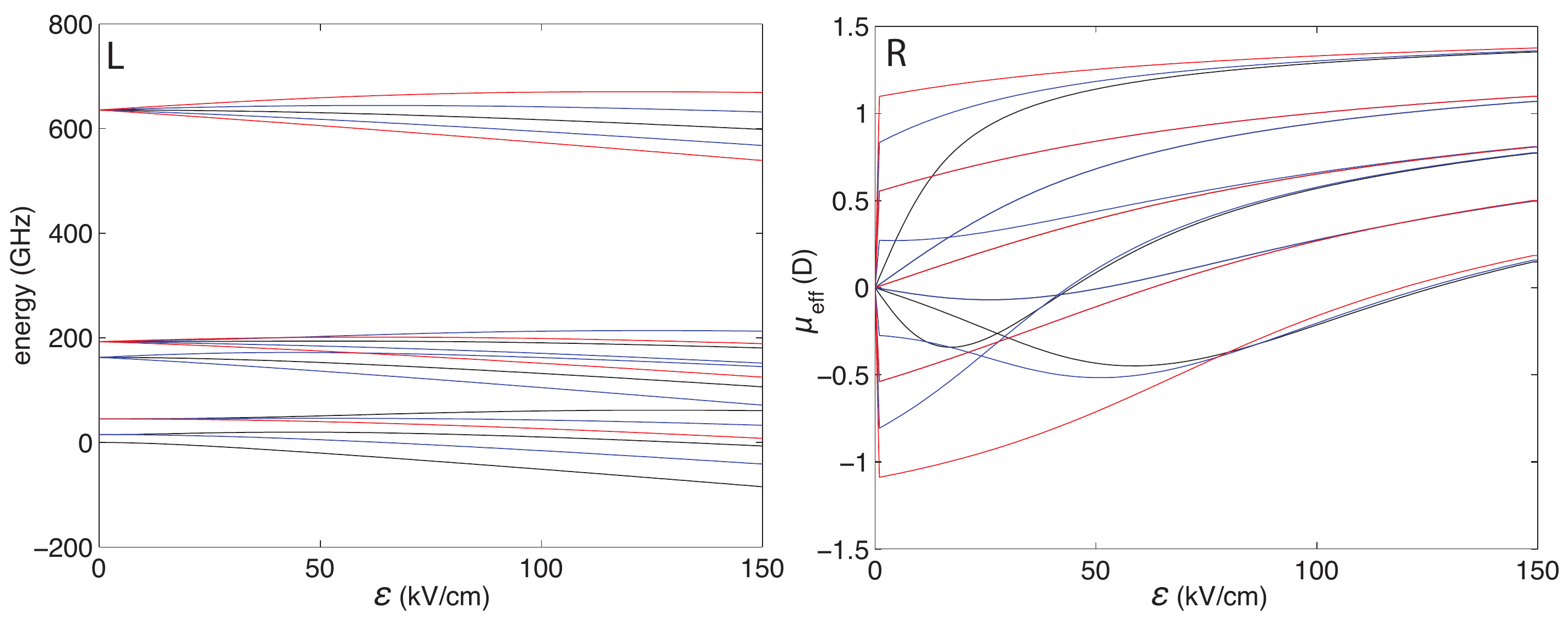}\label{fig:stark-curve-2}}\\
   \subfigure[indole]{\includegraphics[width=\linewidth]{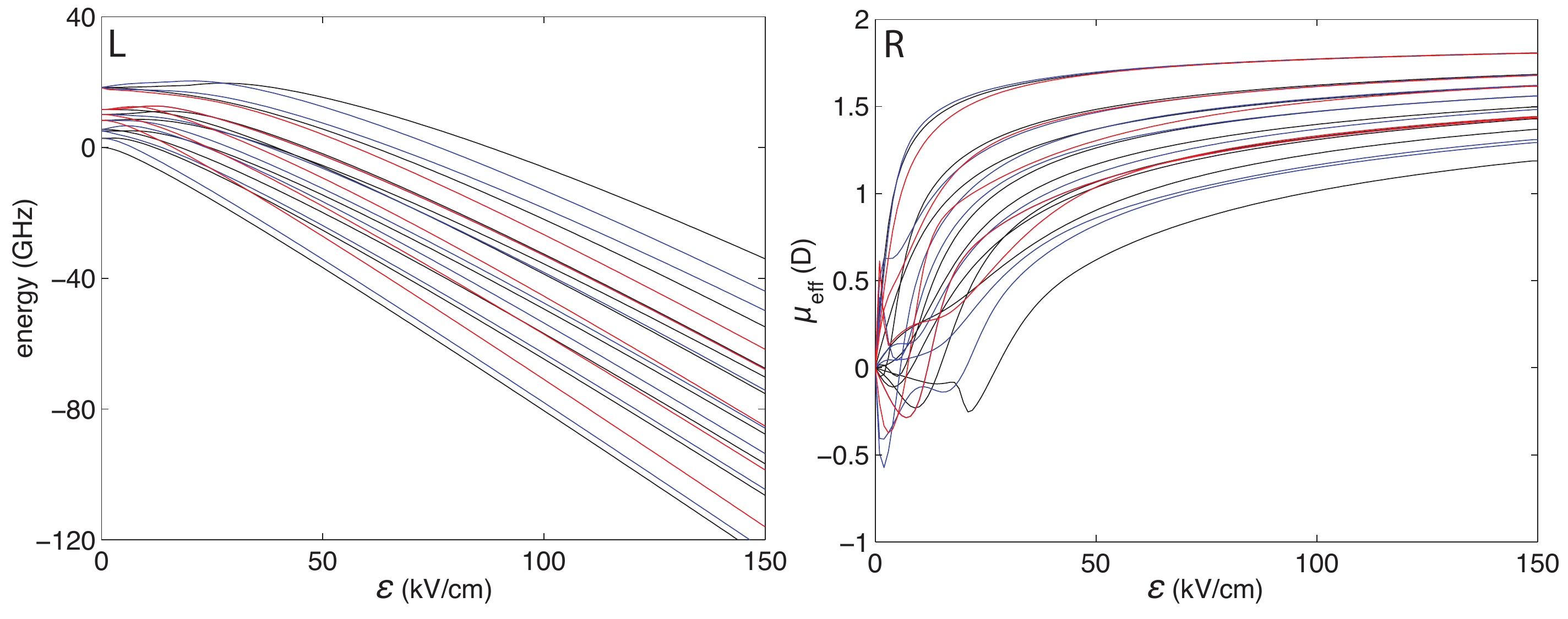}\label{fig:stark-curve-3}}\\
   \caption{(L) Stark energies and (R) effective dipole moments ($\mueff$) of OCS, iodomethane, and
      indole for the $M=0$ (black), $M=1$ (blue), and $M=2$ (red) levels of $J=0-2$.}
\end{figure}

In practice, the calculation of the Stark energies is performed for a number of electric field
strengths -- typically in steps of 1\,kV/cm from 0\,kV/cm to 200\,kV/cm -- and the resulting
energies are stored for later use. The calculated Stark curves and effective dipole moments for
lowest-lying rotational quantum states of OCS (linear rotor), iodomethane (symmetric top), and
indole (asymmetric top) are plotted using \texttt{cmistark\_plot\_energies} and are shown in
\autoref{fig:stark-curve-1}, \autoref{fig:stark-curve-2}, and \autoref{fig:stark-curve-3},
respectively.
Here the effective dipole moment, $\mueff$, is introduced as:
\begin{equation}
   \mueff(\field)=-\frac{\partial E(\field)}{\partial \field}
\end{equation}
This is the space fixed dipole moment, \ie, the projection of the molecular frame dipole moment onto
the field direction. It is extremely useful in further simulations on the manipulation of polar
molecules with inhomogeneous electric fields, where the force exerted on the molecule is directly
proportional to $\mueff$~\cite{Bethlem:JPB39:R263}.

Firstly, in these figures, the Stark energy curves from different $M$ always cross, \ie, $M$ is a
good quantum number. Secondly, while the energy curves of OCS (\autoref{fig:stark-curve-1}~L) and
iodomethane (\autoref{fig:stark-curve-2}~L) show relatively simple structures, those for indole
(\autoref{fig:stark-curve-3}~L) show more complicated behavior. Furthermore, for curves of indole of
each $M$ (\autoref{fig:stark-curve-3}~L), most crossings between Stark curves are avoided, because
the dipole moment of indole is not parallel to any principal axis ($\mu_a\neq0,\mu_b\neq0,\mu_c=0$).
Thus, for $M\neq0$ all states in the field have the same symmetry. For OCS and iodomethane, the sign
of the effective dipole moment, \mueff, can be negative or positive, depending on the quantum state
and field strength, as shown in \autoref{fig:stark-curve-1}~R and \autoref{fig:stark-curve-2}~R.
However, for indole, the sign of \mueff is mostly positive, as shown in
\autoref{fig:stark-curve-3}~R. The rapid changes of signs and values of \mueff shown in
\autoref{fig:stark-curve-3}~R are due to avoided crossings between Stark curves.

In order to evaluate the performance, we have calculated the energy curves for OCS at 151 field
strengths in the range 0 to 150~kV/cm using the different algorithms for three types of rotors for
all states up to $J=32$. This yields the following computation times on a current iMac:
\begin{itemize}
\item{Linear rotor code: 1.0~s}
\item{Symmetric-top code: 55~s}
\item{Asymmetric-top code: 300~s (3~min 20~s, 150~\% CPU utilization)}
\end{itemize}
Note that the runtime largely reflects the time spent on diagonalizing the matrix, and thus the size
of the matrix. According to the LAPACK benchmark report in LAPACK Users'
Guide~\cite{Anderson:Lapack}, for diagonalizing dense symmetric $N$ by $N$ matrices by using dsyevr,
the computing time for $N=1000$ is about 400 times of that for $N=100$. However, the computing time
for $N=2000$ is about 10 times of that for $N=1000$. In practice, for asymmetric top calculations an
increase of the maximum $J$ by $10$ (\eg, $J=40\rightarrow50$ or $J=90\rightarrow100$) included in
the calculation roughly doubles the runtime.

\section{Installation instructions}
\label{sec:installation_ins}

\subsection{Requirements}
\label{sec:requirements}
\cmistark needs an operational Python installation, the external Python packages,
numpy, scipy, PyTables, matplotlib, and a command-line interface to start the various python scripts provided
here.

\subsection{Obtaining the code}
\label{sec:obtaining-code}
The program is available from CPC Program Library, Queen's University, Belfast, N. Ireland. The
latest version of the program can also be obtained from the Controlled Molecule Imaging (CMI) group.

\subsection{Installation}
\label{sec:installation}
Installation is performed by executing the generic Python install command \texttt{python setup.py
   install} in the unpacked source code directory. This requires a write access to the packages
directory tree of the Python distribution. Alternatively, on Unix-like systems the provided
shell-script \texttt{user-install.sh} can be used to install the program into an user-specified
directory, such as \texttt{\$HOME/.python}. This method requires the user to define the shell
variable \texttt{PYTHONHOME} to include this directory in the python search path.

\section{Documentation}
\label{sec:documentation}
\sloppy A full documentation is provided within the source code and only briefly summarized here. To
perform a Stark effect calculation the script file \texttt{cmistark\_calculate\_energy} is used.
Some of its general command-line options are
\begin{itemize}
 \item{\texttt{--<moleculename>}: specify which molecule is used in the calculation,}
 \item{\texttt{--dc-fields}: specify the range of the dc electric field strength,}
 \item{\texttt{--Jmax\_calc}: specify the maximum value of $J$ included in the calculation,}
 \item{\texttt{--Jmax\_save}: specify the maximum value of $J$ of Stark curves saved in the output
       file.}
\end{itemize}
Two scripts \texttt{cmistark\_plot\_energy} and \texttt{cmistark\_print\_energy} are provided to
access existing files with stored Stark curves, and plot or convert to text, respectively, all or
selected Stark energy curves.

Calculating the Stark curves of OCS is as simple as running the command
\begin{verbatim}
cmistark_calculate_energy --Jmax_save=2 --Jmax_calc=10 \
    --dc-fields=0:150:151 --OCS
\end{verbatim}
The data is saved in OCS.molecule and no command-line output is produced. The correct output
resulting from this calculation is provided in the \texttt{samples/} directory of the source code.
The plot in \autoref{fig:stark-curve-1}\,L can then be produced by running
\begin{verbatim}
 cmistark_plot_energy OCS.molecule
\end{verbatim}

Currently, isotopologues of the following molecules are implemented in
\texttt{cmistark\_calculate\_energy}, with parameters from the literature as referenced in the code:
3-aminophenol, carbonylsulfide, water, indole, indole(water)$_1$, iodomethane, difluoroiodobenzene,
aminobenzonitrile, benzonitrile, iodobenzene, and sulfurdioxide. Implementing a new molecule is as
simple as adding a code block in \texttt{cmistark\_calculate\_energy} to define relevant molecular
parameters, molecular constants and dipole moment components. For the \emph{cis} and \emph{trans}
conformers of 3-aminophenol~\cite{Filsinger:PCCP10:666}, this is implemented in the following way:

\medskip
\begin{minipage}{1.0\linewidth}\footnotesize
\begin{verbatim}
def three_aminophenol(param):
    # Inertial parameters: Filsinger et al, Phys. Chem. Chem. Phys. 10, 666 (2008)
    # dipole moment:       Filsinger et al, Phys. Chem. Chem. Phys. 10, 666 (2008)
    param.name = "3-aminophenol"
    param.watson = 'A' # specify Watson's A reduction
    param.symmetry = 'N' # specify "no symmetry" in the dc electric field
    if param.isomer == 0: # cis-conformer
        # the following line specifies the rotational constants (A, B, C)
        # in unit of Hz.
        param.rotcon = convert.Hz2J(numpy.array([3734.93e6, 1823.2095e6, 1226.493e6]))
        # the following line specifies the dipole moment components ($\mu_a,\mu_b,\mu_c$)
        # in unit of Debye.
        param.dipole = convert.D2Cm(numpy.array([1.7718, 1.517, 0.]))
    elif param.isomer == 1: # trans-conformer
        param.rotcon = convert.Hz2J(numpy.array([3730.1676e6, 1828.25774e6, 1228.1948e6]))
        param.dipole = convert.D2Cm(numpy.array([0.5563, 0.5375, 0.]))
\end{verbatim}
\end{minipage}
\medskip

\section{Alternative software}
\label{sec:alternative-software}
Several programs exist for the simulation of rotationally resolved spectra of asymmetric top
molecules, such as \textsc{pgopher}~\cite{Western:pgopher},
\textsc{spfit/spcat}~\cite{Pickett:JMolSpec148:371, spfit},
\textsc{qstark}~\cite{Kisiel:JPC104:6970, Kisiel:CPL325:523, Kisiel:QSTARK},
\textsc{dbsrot}~\cite{Borst:thesis:2001, Kang:JCP122:174301},
\textsc{krot}~\cite{Kuepper:thesis:2000}, \textsc{asyrot}~\cite{Birss1984} and
\textsc{jb95}~\cite{Majewski:LTC}, as well as programs for automated fitting of high resolution
spectra, \eg, based on genetic algorithm~\cite{Meerts:CanJC82:804}. Inherently these programs work
by calculating the energies of all states possibly involved in the relevant transitions, \ie, they
do perform similar calculations as \cmistark. So far, to the best of our knowledge, only the
programs \textsc{pgopher}~\cite{Western:pgopher} and \textsc{qstark}~\cite{Kisiel:JPC104:6970,
   Kisiel:CPL325:523, Kisiel:QSTARK} can also calculate Stark energies of linear, symmetric, and
asymmetric rotors. The program \textsc{qstark}~\cite{Kisiel:JPC104:6970, Kisiel:CPL325:523,
   Kisiel:QSTARK} allows calculations including quadrupole coupling effects for one nucleus.
\textsc{pgopher}~\cite{Western:pgopher} can take into account some internal motions, such as
internal rotations (torsion) or inversion motions, \eg, in NH$_3$. These effects will be implemented
in future versions of \cmistark. However, they require considerably more intricate handling of
symmetry properties. However, the available programs are not well suited for simulations in the
controlled molecules field where it is necessary to calculate Stark energies in very strong fields
and to correctly label large numbers of quantum states over the full field-strength regime. For
example, while the program \textsc{qstark} calculates Stark energies essentially correctly in strong
fields, labeling problems are known when the off-diagonal elements in the $H$ matrix become
sufficiently large~\cite{Kisiel:QSTARK}. \textsc{pgopher} provides direct access to the Stark curves
of individual or a few quantum states. However, its graphical/text based access is not convenient
for the calculation and storage of many precisely calculated Stark curves with a sufficiently large
range of quantum states. Note that, even for relatively small complex molecules, such as
benzonitrile or indole under conditions of only few K, many thousand Stark curves need to be
calculated with $J$ up to 50, with hundreds of energies per curve for specific dc field strengths,
and they must be stored for easy retrieval in further calculations.

\section{Outlook}
\label{sec:outlook}
The current program has been successfully used in the calculation of Stark energy maps of various
asymmetric top molecules, for instance, benzonitrile~\cite{Wohlfart:PRA77:031404},
4-aminobenzonitrile~\cite{Filsinger:PRA82:052513}, 3-aminophenol~\cite{Filsinger:PRL100:133003,
   Filsinger:ACIE48:6900}, indole, and indole-water clusters~\cite{Trippel:PRA86:033202}. Those
calculation results from the progam were successfully applied to fit and analyze experimental data
on the manipulation of molecules with electric fields. The program was also tested against the
energies of lowest rotational states from \textsc{qstark}~\cite{Kisiel:JPC104:6970,
   Kisiel:CPL325:523, Kisiel:QSTARK}, with a relative error on the order of $10^{-6}$ limited by the
numerical precision of slightly different implementations of the Hamiltonian and the matrix
diagonalization.

The current program will be further improved in several directions. For example, for molecules
containing large nuclear quadrupole constants the corresponding quadrupole coupling terms need to be
implemented. The challenge here is to still automatically symmetrize the Hamiltonian and to
correctly label the resulting states. Moreover, especially many of the small molecules employed in
electric-field manipulation experiments are open-shell, \ie, they possess electronic (orbital and
spin) angular momentum. The respective Hamiltonians could also be implemented in \cmistark. We will
implement such extensions as they are relevant for the simulation of our manipulation experiments.
We will support third parties to extend our code to their needs, under the provision that it is
provided to all users after a reasonable amount of time.

\section*{Acknowledgments}
\label{sec:acknowledgments}
We thank Rosario González-Férez, Bas van der Meerakker, Gerard Meijer, and members of the CFEL-CMI
group for helpful discussions. 
Izan Castro Molina implemented the initial linear and symmetric top calculations. This work has been
supported by the DFG priority program 1116 ``Interactions in ultracold and molecular gases'' and by
the excellence cluster ``The Hamburg Center for Ultrafast Imaging -- Structure, Dynamics and Control
of Matter at the Atomic Scale'' of the Deutsche Forschungsgemeinschaft.

\appendix
\section{Matrix elements for linear rotors}
\label{app:lin_rot}
For the linear top, the matrix elements of $H_\text{rigid}$ and $H_\text{d}$ can be written
as~\cite{Gordy:MWMolSpec}:
\begin{equation}
   \langle J,M|H_\text{rigid}|J,M\rangle = hBJ(J+1),
\end{equation}
\begin{equation}
   \langle J,M|H_\text{d}|J,M\rangle = -hDJ^2(J+1)^2,
\end{equation}
where $B$ (Hz) and $D$ (Hz) are the corresponding rotational constant and the quartic centrifugal
distortion constant, respectively. The matrix elements for the Stark Hamiltonian $H_\text{Stark}$
are expressed as following~\cite{Zare:AngularMomentum}:
\begin{align}
   \langle J+1,M|H_\text{Stark}|J,M\rangle
   & = \langle J,M|H_\text{Stark}|J+1,M\rangle \notag \\
   & = -\frac{\sqrt{(J+1)^2-M^2}}{\sqrt{(2J+1)(2J+3)}}\mu\field
\end{align}

\section{Matrix elements for symmetric tops}\label{app:sym_rot}
For the rigid prolate and oblate symmetric top the matrix elements of $H_\text{rigid}$ can be
written as~\cite{Gordy:MWMolSpec}:
\begin{align}
   \langle J,K,M|H_\text{rigid}|J,K,M\rangle &= h\left[BJ(J+1)+(A-B)K^2\right] \quad \text{(prolate)} \\
   \langle J,K,M|H_\text{rigid}|J,K,M\rangle &= h\left[BJ(J+1)+(B-C)K^2\right] \quad \text{(oblate)}
\end{align}
with the rotational constants $A$, $B$, $C$ (Hz). The matrix elements of $H_\text{d}$ are expressed
as following~\cite{Gordy:MWMolSpec}:
\begin{align}
   \langle J,K,M|H_\text{d}|J,K,M\rangle =-h\left[\Delta_JJ^2(J+1)^2+\Delta_{JK}J(J+1)K^2+\Delta_KK^4\right]
\end{align}
where $\Delta_J, \Delta_{JK}$ and $\Delta_K$ are the first-order (quartic) centrifugal distortion
constants (Hz). The matrix elements of $H_\text{Stark}$ are~\cite{Zare:AngularMomentum}:
\begin{align}
   \langle J,K,M|H_\text{Stark}|J,K,M\rangle &= -\frac{MK}{J(J+1)}\mu\field \\[1ex]
   \langle J+1,K,M|H_\text{Stark}|J,K,M\rangle &=\langle J,K,M|H_\text{Stark}|J+1,K,M\rangle \nonumber \\
   &= -\frac{\sqrt{(J+1)^2-K^2}\sqrt{(J+1)^2-M^2}}{(J+1)\sqrt{(2J+1)(2J+3)}}\mu\field
\end{align}

\section{Matrix elements for asymmetric tops}
\label{app:asym_rot}
For the rigid asymmetric top, the matrix elements of $H_\text{rigid}$ in terms of $I^r$
representation~\cite{Gordy:MWMolSpec} can be written as~\cite{Gordy:MWMolSpec,
   Zare:AngularMomentum}:
\begin{equation}
   \langle J,K,M|H_\text{rigid}|J,K,M\rangle=h\left[\frac{B+C}{2}(J(J+1)-K^2)+AK^2\right],
\end{equation}
\begin{align}
   &\langle J,K+2,M|H_\text{rigid}|J,K,M\rangle=\langle J,K,M|H_\text{rigid}|J,K+2,M\rangle \nonumber \\
   &=\frac{h(B-C)}{4}\sqrt{J(J+1)- K(K+1)}\sqrt{J(J+1)-(K+1)(K+2)},
\end{align}
with the rotational constants $A$, $B$, $C$ (Hz). The distortable rotor is described using Watson's
A reduction~\cite{Watson:VibSpecStruct6:1}:
\begin{align}
   \langle J,K,M|H_\text{d}|J,K,M\rangle =-h\left[\Delta_J(J(J+1))^2+\Delta_{JK}J(J+1)K^2+\Delta_KK^4\right],
\end{align}
\begin{align}
   \langle J,K+2,M|H_\text{d}|J,K,M\rangle =& \;\langle J,K,M|H_\text{d}|J,K+2,M\rangle \nonumber\\
   =& \;-h\left[\delta_JJ(J+1)+\frac{\delta_K}{2}((K+2)^2+K^2)\right] \nonumber\\
   & \times\sqrt{J(J+1)-K(K+1)} \\
   & \times\sqrt{J(J+1)-(K+1)(K+2)} \nonumber
\end{align}
with the five linearly independent quartic distortion constants $\Delta_J, \Delta_{JK}, \Delta_K,
\delta_J$ and $\delta_K$ (Hz). The contribution of $\mu_a$, \ie, the dipole moment component along
the principal axis of inertia $a$, is~\cite{Zare:AngularMomentum,Cross:JCP12:210}:
\begin{equation}
\langle J,K,M|H^{a}_\text{Stark}|J,K,M\rangle=-\frac{MK}{J(J+1)}\mu_a\field
\label{eq:asy_stark_a_diag}
\end{equation}
\begin{align}
   \langle J+1,K,M|H^{a}_\text{Stark}|J,K,M\rangle &=\langle J,K,M|H^{a}_\text{Stark}|J+1,K,M\rangle\nonumber \\
   &= -\frac{\sqrt{(J+1)^2-K^2}\sqrt{(J+1)^2-M^2}}{(J+1)\sqrt{(2J+1)(2J+3)}}\mu_a\field
\label{eq:asy_stark_a_off}
\end{align}
The contribution of $\mu_b$ is:
\begin{align}
&\langle J,K+1,M|H^{b}_\text{Stark}|J,K,M\rangle=-\frac{M\sqrt{(J-K)(J+K+1)}}{2J(J+1)}\mu_b\field
\label{eq:asy_stark_b_diag}
\end{align}
\begin{multline}
   \langle J+1,K\pm1,M|H^{b}_\text{Stark}|J,K,M\rangle \\
   =\pm\frac{\sqrt{(J\pm K+1)(J\pm K+2)}\sqrt{(J+1)^2-M^2}}{2(J+1)\sqrt{(2J+1)(2J+3)}}\mu_b\field
\end{multline}
The $H_\text{Stark}$ matrix elements involving $\mu_c$ are:
\begin{align}
&\langle J,K\pm 1,M|H^{c}_\text{Stark}|J,K,M\rangle=\pm i\frac{M\sqrt{(J\mp K)(J\pm K+1)}}{2J(J+1)}\mu_c\field
\label{eq:asy_stark_c_diag}
\end{align}
\begin{multline}
   \langle J+1,K\pm1,M|H^{c}_\text{Stark}|J,K,M\rangle \\
   =-i\frac{\sqrt{(J\pm K+1)(J\pm K+2)}\sqrt{(J+1)^2-M^2}}{2(J+1)\sqrt{(2J+1)(2J+3)}}\mu_c\field
\end{multline}
Note that the equations above use the representation $I^r$ with the phase convention and formalism
of Zare~\cite{Zare:AngularMomentum}.

\section{Fourgroup}
\label{app:four_group}
The symmetry properties of the rotational Hamiltonian, as well as rotational wavefunctions, of a
rigid asymmetric top molecule may be deduced from its ellipsoid of inertia, which is symmetric not
only to an identity operation $E$ but also to a rotation by $180\degree$, a $C_2$ operation, about
any of its principal axes of inertia. This set of symmetry operations forms the Fourgroup
(\emph{Viergruppe}), which is designated by $V(a,b,c)$~\cite{Gordy:MWMolSpec}. These symmetry
operations cause the angular momentum to transform in the following manner~\cite{Gordy:MWMolSpec}:
\begin{align}
   E:     &\quad \op{J}_a\rightarrow{\op{J}_a},\op{J}_b\rightarrow{\op{J}_b},\op{J}_c\rightarrow{\op{J}_c} \\
   C^a_2: &\quad \op{J}_a\rightarrow{\op{J}_a},\op{J}_b\rightarrow{-\op{J}_b},\op{J}_c\rightarrow{-\op{J}_c} \\
   C^b_2: &\quad \op{J}_a\rightarrow{-\op{J}_a},\op{J}_b\rightarrow{\op{J}_b},\op{J}_c\rightarrow{-\op{J}_c} \\
   C^c_2: &\quad \op{J}_a\rightarrow{-\op{J}_a},\op{J}_b\rightarrow{-\op{J}_b},\op{J}_c\rightarrow{\op{J}_c}
\end{align}
The character table of the Fourgroup can is shown in \autoref{tab:fourgroupv:character-table}.
\begin{table}[b]
   \begin{center}
      \begin{tabular}{ccccc}
         \hline\hline
         V & E & $C_2^a$ & $C_2^b$  & $C_2^c$ \\
         \hline
         A & 1 & 1 & 1 & 1 \\
         B$_a$ & 1 & 1 & -1 & -1 \\
         B$_b$ & 1 & -1 & 1 & -1 \\
         B$_c$ & 1 & -1 & -1 & 1 \\
         \hline\hline
      \end{tabular}
      \caption{character table for the four group V}
      \label{tab:fourgroupv:character-table}
   \end{center}
\end{table}

\section{Wang transformation}
\label{app:Wang_trans}
The field-free semirigid rotor Hamiltonian operators $H_\text{rigid}+H_\text{d}$ described above can
be symmetrized to belong to the Fourgroup V and every field-free rotor wavefunction can be
classified according to its behavior under $V(a,b,c)$~\cite{Gordy:MWMolSpec}. This symmetry
classification is provided in \autoref{tab:Wang:symmetry}.
\begin{table}[b]
   \begin{center}
      \begin{tabular}{ccccc}
         \hline\hline
         submatrix & $K$ & $s$ & $J_{even}$  & $J_{odd}$ \\
         \hline
         $E^+$ & e & 0 & $A(ee)$   & $B_a(eo)$ \\
         $E^-$ & e & 1 & $B_a(eo)$ & $A(ee)$ \\
         $O^+$ & o & 0 & $B_b(oo)$ & $B_c(oe)$ \\
         $O^-$ & o & 1 & $B_c(oe)$ & $B_b(oo)$ \\
         \hline\hline
      \end{tabular}
      \caption{Symmetry classification of asymmetric top wavefunctions $|J_{K_aK_c},M\rangle$ for
         representation $I^r$~\cite{Zare:AngularMomentum}. The symmetry species of each $J_{K_aK_c}$
         is determined by the eveness or oddness of $K_a$ and $K_c$, which is indicated in
         parentheses in columns 4 and 5. The classification of Wang sub-matrices is also provided.}
      \label{tab:Wang:symmetry}
   \end{center}
\end{table}

The symmetrized basis functions constructed by Wang transformation are defined
as~\cite{Gordy:MWMolSpec, Wang:PR34:243, Mulliken:PR59:873}:
\begin{align}
   |J,0,M,0\rangle &= |J,0,M\rangle &&\quad\text{for } K=0 \\
   |J,K,M,s\rangle &= \frac{1}{\sqrt{2}}(|J,K,M\rangle+(-1)^s|J,-K,M\rangle) &&\quad\text{for } K\neq0
\end{align}
where $s$ is 0 (symmetric) or 1 (antisymmetric) and $K$ now takes on only positive values. The Wang
transformation can be expressed in a matrix form and the transformation matrix $X$ of order $(2J+1)$
can be expressed as:
\begin{equation}
   X = X^{-1} = \tilde{X} = \frac{1}{\sqrt{2}}\begin{bmatrix}
      \ddots &    &    & \vdots   &   &   & \reflectbox{$\ddots$} \\
      & -1 &  0 & 0        & 0 & 1 &        \\
      &  0 & -1 & 0        & 1 & 0 &        \\
      \cdots &  0 &  0 & \sqrt{2} & 0 & 0 & \cdots \\
      &  0 &  1 & 0        & 1 & 0 &        \\
      &  1 &  0 & 0        & 0 & 1 &        \\
      \reflectbox{$\ddots$} &    &    & \vdots   &   &   & \ddots \\
   \end{bmatrix}
\end{equation}
The change of basis can be written as $\mathbf{\Psi}_\text{Wang}=\tilde{X}\mathbf{\Psi}$. For fixed
values of $J$ and $M$, the vector $\mathbf{\Psi}$ consists of $(2J+1)$ symmetric top basis functions
$|J,K,M\rangle$, whereas $\mathbf{\Psi}_\text{Wang}$ is the vector of new basis functions that
contains the $(2J+1)$ symmetrized functions $|J,K,M,s\rangle$:
\begin{equation}
   \mathbf{\Psi}_\text{Wang} =\begin{pmatrix}
      |J,J,M,1\rangle \\
      |J,(J-1),M,1\rangle \\
      \vdots \\
      |J,1,M,1\rangle \\
      |J,0,M,0\rangle \\
      |J,1,M,0\rangle \\
      \vdots \\
      |J,(J-1),M,0\rangle \\
      |J,J,M,0\rangle \\
   \end{pmatrix}, \quad
   \mathbf{\Psi} =\begin{pmatrix}
      |J,-J,M\rangle \\
      |J,(-J+1),M\rangle \\
      \vdots \\
      |J,-1,M\rangle \\
      |J,0,M\rangle \\
      |J,1,M\rangle \\
      \vdots \\
      |J,(J-1),M\rangle \\
      |J,J,M\rangle \\
   \end{pmatrix}
\end{equation}
In the new basis the Hamiltonian matrix factorizes into four sub-matrices that are historically
denoted as $E^+,O^+,E^-,O^-$~\cite{Wang:PR34:243, Mulliken:PR59:873}:
\begin{equation}
   H_\text{Wang,\,rot}=\tilde{X}H_\text{rot}X=E^+ + O^+ + E^- + O^-
   \label{eq:Wang:ks}
\end{equation}
These sub-matrices are classified by the eveness and oddness of $K$ and $s$, as shown in
\autoref{tab:Wang:symmetry}. For a single value of $J$, $|J,K,M,s\rangle$ wavefunctions within each
sub-matrix all belong to a same symmetry species of $V$~\cite{Hainer:JCP17:826, Gordy:MWMolSpec} and
the correlation is given in \autoref{tab:Wang:symmetry}. Thus, $H_\text{Wang,\,rot}$ can also be
block-diagonalized in terms of four symmetry species, $A$, $B_a$, $B_b$, and $B_c$. This
symmetrization of the basis by the Wang transformation simplifying the numerical evaluation and,
most importantly, is necessary for the correct adiabatic labeling of the eigenstates in the electric
field.

\section{Block diagonalization of the Hamiltonian matrix}
\label{app:sym_field}
The Hamiltonian is block-diagonal in $M$ and the calculations are performed for each $M$ separately.
In the field-free case all $M$s are degenerate and only $M=0$ is calculated. For the symmetric top,
$K$ is a good quantum number and the matrix is always also factorized into separate $K$ blocks. As
mentioned in \ref{app:Wang_trans}, the Hamiltonian matrix of field-free symmetric or asymmetric tops
can be block diagonalized into four blocks according to Fourgroup symmetry. An external dc electric
field can mix these blocks, but remaining symmetries allow partial factorization. In
\autoref{tab:symmetries} we summarize the block diagonalization of the Hamiltonian matrix in an
electric field according to $V$ for all possible cases of non-zero dipole moment directions, \ie,
all possible combinations of non-zero dipole-moment components in the principal axes of inertia
system. We note that the remaining symmetry can be higher for $M=0$ than for $M\neq0$. This can also
be seen from the matrix elements given above, where the $\Delta{J}=0$ Stark-coupling elements are
always proportional to $M$, \ie, these couplings vanish for $M=0$.
\begin{table}[b]
   \centering\renewcommand{\arraystretch}{1.5}%
   \begin{tabular}{cccp{0.01cm}ccccccccc}
      \hline\hline
      \multicolumn{3}{c}{} & & \multicolumn{4}{ c }{$J_{even}$} & & \multicolumn{4}{ c }{$J_{odd}$} \\
      \cline{5-8} \cline{10-13}
      \multicolumn{3}{c}{} & & $A$ & $B_a$ & $B_b$ & $B_c$ & & $A$ & $B_a$ & $B_b$ & $B_c$ \\
      \cline{5-8} \cline{10-13}
      $\mu_a$ & $\mu_b$ & $\mu_c$ & & $E^+$ & $E^-$ & $O^+$ & $O^-$ & & $E^-$ & $E^+$ & $O^-$ & $O^+$ \\ \hline
      $\neq 0$ & $0$ & $0$ & & $\square \blacksquare$ & $\bigcirc \blacksquare$ & $\triangle \blacklozenge$ & $\Diamond \blacklozenge$ & & $\bigcirc \blacksquare$ & $\square \blacksquare$ & $\Diamond \blacklozenge$ & $\triangle \blacklozenge$ \\
      $0$ & $\neq 0$ & $0$ & & $\square \blacksquare$ & $\bigcirc \blacklozenge$ & $\triangle \blacksquare$ & $\Diamond \blacklozenge$ & & $\triangle \blacksquare$ & $\Diamond \blacklozenge$ & $\square \blacksquare$ & $\bigcirc \blacklozenge$ \\
      $0$ & $0$ & $\neq 0$ & & $\square \blacksquare$ & $\bigcirc \blacklozenge$ & $\triangle \blacklozenge$ & $\Diamond \blacksquare$ & & $\lozenge \blacksquare$ & $\triangle \blacklozenge$ & $\bigcirc \blacklozenge$ & $\square \blacksquare$  \\
      $\neq 0$ & $\neq 0$ & $0$ & & $\square \blacksquare$ & $\lozenge \blacksquare$ & $\lozenge \blacksquare$ & $\square \blacksquare$ & & $\lozenge \blacksquare$ & $\square \blacksquare$ & $\square \blacksquare$ & $\lozenge \blacksquare$ \\
      $0$ & $\neq 0$ & $\neq 0$ & & $\square \blacksquare$ & $\square \blacksquare$ & $\lozenge \blacksquare$ & $\lozenge \blacksquare$ & & $\lozenge \blacksquare$ & $\lozenge \blacksquare$ & $\square \blacksquare$ & $\square \blacksquare$  \\
      $\neq 0$ & $0$ & $\neq 0$ & & $\square \blacksquare$ & $\lozenge \blacksquare$ & $\square \blacksquare$ & $\lozenge \blacksquare$ & & $\lozenge \blacksquare$ & $\square \blacksquare$ & $\lozenge \blacksquare$ & $\square \blacksquare$  \\
      $\neq 0$ & $\neq 0$ & $\neq 0$ & & $\square \blacksquare$ & $\square \blacksquare$ & $\square \blacksquare$ & $\square \blacksquare$ & & $\square \blacksquare$ & $\square \blacksquare$ & $\square \blacksquare$ & $\square \blacksquare$  \\ \hline\hline
   \end{tabular}
   \caption{Symmetries of asymmetric tops in dc electric fields. Different shapes
      represent the distinct symmetry species (matrix blocks) for the case of $M=0$ (open symbols)
      and $M\neq0$ (filled symbols)}
   \label{tab:symmetries}
\end{table}

The factorization summarized in \autoref{tab:symmetries} can be understood in terms of the symmetry
properties of the direction cosine $\phi_{Z_g}$ in \eqref{eq:Hstark}~\cite{Gordy:MWMolSpec}. For the
case of $\mu=\mu_\alpha$ and $M\neq0$, basis functions of symmetries $A$ and $B_\alpha$ are coupled,
as well as those of symmetries $B_{\alpha'}$ and $B_{\alpha''}$, where
$\alpha\neq\alpha'\neq\alpha''\neq\alpha$. However, no coupling between these two subsets exist. The
Hamiltonian matrix can thus be factorized into two blocks, as listed in \autoref{tab:symmetries},
one (filled square symbol) containing $A$ and $B_\alpha$ and the other one (filled diamond symbol)
containing $B_{\alpha'}$ and $B_{\alpha''}$. For the special case of $M=0$, states of symmetries $A$
and $B_\alpha$ for any given $J$ are also not coupled, nor are states of symmetries $B_{\alpha'}$
and $B_{\alpha''}$ coupled~\cite{Escribano:PRA62:023407}. This is due to the vanishing matrix
elements \eqref{eq:asy_stark_a_diag}, \eqref{eq:asy_stark_b_diag}, and \eqref{eq:asy_stark_c_diag}
for $M=0$. As a result, states of symmetry $A$ in $J_{even}$ ($J_{odd}$) only couple with those of
symmetry $B_\alpha$ in $J_{odd}$ ($J_{even}$), etc. This remaining symmetry for the case
$\mu=\mu_\alpha$ is represented by the open square symbol (open circle symbol) in the first line of
\autoref{tab:symmetries}. States of symmetries $B_{\alpha'}$ and $B_{\alpha''}$ also couple in the
same manner, and their remaining symmetries are represented by open triangle and open diamond
symbols in \autoref{tab:symmetries}. Similar behavior is observed for all cases when the dipole
moment is along a principal axis of inertia. For the cases with more than one non-zero dipole moment
component states of all four symmetry species $A$, $B_a$, $B_b$ and $B_c$ are coupled for $M\neq0$
and only one symmetry species remains. For $M=0$ and one $\mu_\alpha=0$, the Hamiltonian matrix can
be factorized into two blocks. For a dipole moment with components along all principal axes of
inertia no partial Fourgroup symmetry remains in an electric field.

\section*{References}
\label{sec:references}
\bibliographystyle{model1a-num-names}
\bibliography{string,cmi}
\end{document}